\documentclass[pre,twocolumn,a4paper,amsmath,amssymb,
floatfix,superscriptaddress]{revtex4-1}

\usepackage{graphicx}
\usepackage{epstopdf}

\begin{document}

\title{Structural correlations and melting of B-DNA fibres}

\author{Andrew Wildes}
\affiliation{Institut Laue Langevin,BP 156, 6, rue Jules Horowitz
38042 Grenoble Cedex 9, France}

\author{Nikos Theodorakopoulos}
\affiliation{Theoretical and Physical Chemistry Institute, National Hellenic 
Research Foundation, Vasileos Constantinou 48, 116 35 Athens, Greece}
\affiliation{Fachbereich Physik, Universit\"at Konstanz, 78457
  Konstanz, Germany} 

\author{Jessica Valle-Orero}
\affiliation{Institut Laue Langevin,BP 156, 6, rue Jules Horowitz
38042 Grenoble Cedex 9, France}
\affiliation{Universit\'e de Lyon, Ecole 
Normale Sup\'erieure de Lyon, Laboratoire de Physique CNRS UMR 5672,
46 all\'ee d'Italie, 69364 Lyon Cedex 7, France}

\author{Santiago Cuesta-L\'opez}
\affiliation{Universit\'e de Lyon, Ecole 
Normale Sup\'erieure de Lyon, Laboratoire de Physique CNRS UMR 5672,
46 all\'ee d'Italie, 69364 Lyon Cedex 7, France}

\author{Jean-Luc Garden}
\affiliation{Institut N\'eel, CNRS - Universit\'e Joseph Fourier,
25 rue des Martyrs, BP 166, 38042 Grenoble cedex 9, France.}

\author{Michel Peyrard}
\affiliation{Universit\'e de Lyon, Ecole 
Normale Sup\'erieure de Lyon, Laboratoire de Physique CNRS UMR 5672,
46 all\'ee d'Italie, 69364 Lyon Cedex 7, France}

\date{\today}

\begin{abstract}
Despite numerous attempts, the understanding of the thermal
denaturation of DNA is still a challenge due to the lack of structural
data at the transition since standard experimental approaches to DNA
melting are made in solution and do not provide spatial
information. We report a measurement using neutron scattering from
oriented DNA fibres to determine the size of the regions that stay in
the double-helix conformation as the melting temperature is approached
from below. A Bragg peak from the B-form of DNA has been observed as a
function of temperature and its width and integrated intensity have
bean measured. These results, complemented by a differential
calorimetry study of the melting of B DNA fibres as well as
electrophoresis and optical observation data, are 
analysed in terms of a one-dimensional mesoscopic model of DNA.
\end{abstract}

\pacs{}

\maketitle

\section{Introduction}
\label{sec:intro}

The X-Ray diagrams published by Wilkins et al.\ \cite{Wilkins}
and Franklin et al.\ \cite{Franklin} 
in the same issue of Nature as the famous paper of Watson and
Crick \cite{WatsonCrick} describing the structure of DNA revealed the
significance of the fibre form of DNA in providing oriented samples necessary
for structural studies. These images, showing the cross pattern typical of
a helix and two strong spots associated to the stacking of the bases
in B-DNA, were however illustrating only one aspect of the molecule,
its average static structure. In reality the DNA molecule is a highly
dynamical object. Its base pairs fluctuate widely. The lifetime of a
closed base pair is only of the order of 10 ms \cite{Gueron}. The
local opening of the pairs is important for biological function as it
allows the reading of the genetic code. When temperature is raised
above the physiological range, thermally induced base-pair openings
become more cooperative, leading to the so-called ``DNA bubbles'' i.e.\
open regions which may extend over tens of base pairs. At
sufficiently high temperature they extend over the full molecule and the
two strands separate from each other. For a physicist the thermal
denaturing of DNA, also known as DNA ``melting'', is a phase transition,
which is particularly interesting because it occurs in an
essentially one-dimensional system. DNA melting started to attract
attention soon after the 
discovery of the double helix structure \cite{Thomas,Rice} and was
widely studied, providing
insights on the interactions within DNA, the influence of base pair
sequence on DNA unwinding, and the effect of the solvent on DNA
stability \cite{Wartell}.  It recently attracted a renewed interest
thanks to High Resolution Melting methods (HRM) \cite{RefHRM}
whereby precise denaturing profiles provide a new tool for
biology laboratories. 

\medskip

Despite numerous attempts, the understanding of this remarkable
thermodynamic phase transition is still a theoretical challenge. 
Statistical physics of DNA thermal denaturing has a long history
\cite{Wartell} because it raises the fundamental question of a phase
transition in a one-dimensional system, but also for practical
applications such as the design of Polymerase Chain Reaction (PCR) 
probes or the HRM studies for biology.
The models for DNA denaturation fall in two classes. First, Ising
models treat a base pair as a two-state system, which is either closed
or open. This is the case of the prototype Poland-Scheraga model
\cite{Poland}. Those models are appealing for their simplicity and
because their parameters have been well calibrated. However their
drawback is that they need a large number of phenomenological
parameters and, for genomic sequences, the calculation can become heavy
due to non-local entropy contributions.  The second class of models goes
beyond a description in terms of two-state systems by incorporating
some elements of the structure. The Peyrard-Bishop-Dauxois (PBD)
model \cite{PBD} is still
simple because it represents the status of a base pair by a single
real number measuring the stretching of the bond between the bases,
but contains nevertheless a minimal structural information
relevant for structure factor calculations.
In its simplest version this model allows a fast calculation of melting
curves of long natural DNAs with only 7 parameters \cite{NTh}. 
However the success of different models in describing complex melting
profiles \cite{NTh,JOST} shows that the correct fit of those curves is
not a sufficient test to validate a theory. 
Examination of further observables, with a more direct link to the
underlying structural details, appears necessary.
As for other classes of
phase transitions in physics, such as magnetic systems, 
an important feature that characterises
the nature of a transition is the growth of the size of the
correlated domains as the transition is approached \cite{Cowley}. 

\medskip
Traditional methods to investigate the DNA melting transition 
cannot provide
this kind of spatial information. The standard experimental method is
to record the sharp increase of UV absorbance at 260 nm which is
associated with the un-stacking of the base pairs, while slowly
heating a dilute DNA solution. Other approaches rely on circular
dichroism measurements or calorimetric studies that measure the heat
absorbed at the transition. Although melting curves, showing the
fraction of open base pairs versus temperature, may exhibit a
multi-step behaviour related to the sequence, none of the experiments
are actually sensitive to spatial information, such as the size of the
intact regions of the double helix. 

This structural information, which
is essential to understand the nature of a phase transition, has been
lacking. Neutron scattering can provide this missing piece of
information, provided the experiment can be performed on an oriented
DNA sample. As shown by the historical studies that revealed the
double-helical structure \cite{Wilkins,Franklin} this is possible with
fibre diffraction. The methods to produce fibre samples have been
refined, by controlling the ionic and water content, it is now
possible to make high quality fibres  with various configurational
structures \cite{Ruprecht, Fuller}.

\medskip
 Here we report a measurement using neutron
scattering from oriented DNA fibres (Sec.~\ref{sec:experimental}), 
to determine the size of the
regions that stay in the double-helix conformation as the melting
temperature is approached from below and we show how it can be
analysed in terms of the one-dimensional mesoscopic PBD model of DNA
\cite{PBD} (Secs \ref{sec:theory} and \ref{sec:discussion}).  
We recently published a brief report of those results
\cite{PRL}, which are here presented with further data and discussion.

\section{Experiments}
\label{sec:experimental}

Due to the regular stacking of the base pairs, the B-form of DNA can be viewed
as a one-dimensional diffraction grating. This is reflected in a strong Bragg
peak for a longitudinal component of the
scattering vector $Q_{\parallel} \approx 1.87\;$\AA$^{-1}$, associated 
with the average distance $a = 3.36\;$\AA\ between consecutive base pairs. 
The principle of our experiment is simple:
by following the evolution of this peak as temperature is raised from room
temperature to the denaturation temperature $T_c$, 
we can monitor the breaking of
this ``diffraction grating'' into pieces separated by denatured regions,
where the base stacking is destroyed. We expect a strong broadening of the
diffraction peak as $T_c$ is approached. The width of the peak allows us
to determine how the average size of the
intact double helical domains
evolves when DNA approaches its denaturation temperature, which is
critical information for a theoretical analysis of the transition. 
The interest of this
method that focuses on a single, intense, diffraction peak is that we
precisely collect the information of interest in a 
measurement which is only weakly perturbed by sample imperfections. In
fibres, the B-form of DNA is semicrystalline \cite{Fuller}.
The misalignment of the DNA molecules has been estimated
to be less than 5 degrees \cite{Grimm}. Its effect on the projection of
the base pair distance on the axis is less than $0.5\;$\%, i.e. negligible
with respect to other effects such as the variation of the 
inter base pair distance as a
function of the sequence. As a result, for a cut in reciprocal space
along the fibre axis, the width of the
Bragg peak is not affected. Moreover, by performing a scan off-centre,
i.e.\ with a scattering vector ${\mathbf{Q}}$ which has a non-zero
component orthogonal to the molecular axis, we can also probe the
displacement of the base pairs in the transverse direction. Such a
scan is not immune from the influence of the misalignment of the
molecules which broaden the peak, but it provides interesting data on
the fluctuations due to the opening of the base pairs in the vicinity
of the thermal denaturation. 

\subsection{Materials and methods}
\label{subseq:expmethods}

{\em Sample preparation}

The samples were made from natural DNA extracted from salmon
testes (Fluka). The DNA had been oriented using a ``spinning'' technique
\cite{Ruprecht}, whereby the DNA is precipitated out of a 0.4 M lithium salt
solution, drawn to a fibre and then wound around a bobbin to make a
film of parallel fibres. The samples are then dried, cut from the
bobbin, and then stored for a number of weeks in an atmosphere
humidified to 75\% using $^2$H$_2$O.  This fixed the water content of
the DNA, ensuring a B-form configurational structure and significantly
reduced incoherent neutron scattering from  
protonated hydrogen in the sample.  The B-form was confirmed using
x-ray and neutron diffraction. 

The neutron scattering samples were folded in concertina fashion, thus
preserving the fibre axis direction.  The samples were then 
placed in a niobium envelope and sealed between aluminum plates using
lead wire for the seal. 
The sample cassette was therefore airtight which maintained the water
content throughout the experiment.  The sample mass was $\sim 1$~g. 

{\em Neutron scattering}

\begin{figure}
\includegraphics[width=7cm]{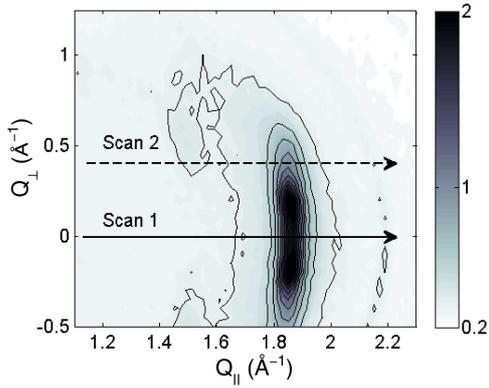}
\caption{Reciprocal space map of B-form Li-DNA. The axes are the
  momentum transfer parallel ($Q_{\|}$) and perpendicular
  ($Q_{\perp}$) to the fibre axis.  The strong Bragg peak is observed
  at $Q_{\|} = 2\pi/a$ where $a \approx 3.36$ {\AA} is the average
  distance between the base pairs along the fibre axis.  A powder
  diffraction peak, coming from the lead wire used to seal the
  cassette, is just visible at $Q = 2.2$ {\AA}.  Also marked on the
  figure are the two standard scans that were repeated at all
  temperatures.
The figure is identical to Fig.~2 \cite{PRL}.} 
\label{fig.Rmap}
\end{figure}

Preliminary diffraction measurements to establish the configurational
form were carried out using the IN3 three axis spectrometer at the
Institut Laue-Langevin, France.  The instrument was configured with a
pyrolytic graphite (PG) monochromator and analyser, and the wavelength
was set to 2.36 {\AA} ($\equiv 14.7$ meV).  The Q-resolution was
defined by $60^{\prime}$ collimation before and after the sample, and
higher order wavelength contamination was suppressed using a PG
filter.  The instrument was used to measure reciprocal space maps, and
an example is shown in figure \ref{fig.Rmap} . 

The neutron three axis spectrometer IN8, also at ILL, was used to
measure the main Bragg peak as a function of temperature.  This
instrument was configured with a PG monochromator delivering an
incident wavelength of 1.53 {\AA} ( $\equiv 35$ meV).  The
Q-resolution was defined with 40$^{\prime}$ collimation before and
after the sample, and was measured by making a reciprocal space map of
the $(111)$ Bragg peak from a silicon single crystal.  No energy
analysis was used,  and the static approximation was assumed to hold
for the measurements.  Temperature control was achieved using a liquid
helium cryofurnace. 

\begin{figure}
\includegraphics[width=7cm]{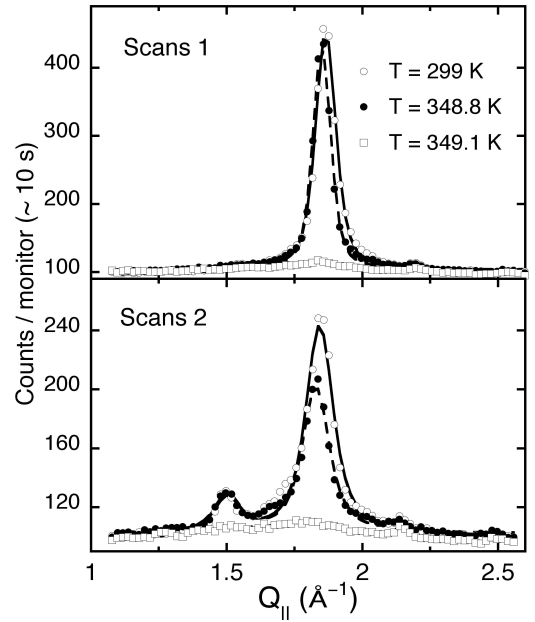}
\caption{Examples of the scans shown in figure \ref{fig.Rmap}.  The
  data at 299 K represent the starting point for the experiment.  The
  sample is in the melting transition at 348.8 K.  The fibre structure
  has collapsed at 349.1 K.  The data have been fitted with equation
  \ref{eq:lorentz} and the fits are also shown.
The figure is identical to Fig.~2 \cite{PRL}.
} 
\label{fig.Scans}
\end{figure}

Two scans were repeated at all temperatures.  Their trajectories were
calculated for nominally elastic scattering and are shown in figure
\ref{fig.Rmap}.  Scan 1 was along the fibre axis, through the centre
of the Bragg peak.  The $Q_{\perp}$ for scan 2 was chosen such that,
when $Q_{\|} = 2\pi/a$, the direction of the scattered beam would be
perpendicular to the fibre axis.  This type of scan has been used to
measure critical phase transitions in low dimensional magnets and
assists the static approximation \cite{Cowley}.  The temperature steps
close to the melting transition were very small (0.1 K) and
measurements at a given temperature were repeated numerous times to
ensure thermal equilibrium and reproducibility.  Examples of the scans
at different temperatures are shown in figure \ref{fig.Scans}.

The data were fitted with the Lorentzian function
\begin{equation}
\label{eq:lorentz}
  S(Q_{\|}) =
  \int^{\infty}_{-\infty}S\left(Q_{\|},\omega\right){\text d}\omega =
  \dfrac{I_0}{\pi} \; \dfrac{ \Gamma/2}{ (\Gamma/2)^2 + (Q_{\|} 
    -Q_0)^2} \; ,
\end{equation}
where $Q_0$ is the peak centre, 
$I_0$ is the integrated intensity and $\Gamma$ is the
width. The function was convoluted with the instrument resolution.  A
second Lorentzian centred at $Q_{\|} \approx 1.5$ {\AA}$^{-1}$ was
needed to fit the scan 2 data.  The amplitudes for these two peaks
were free parameters, however their widths were set to be equal in the
fits.  Examples of the fits are also shown in figure \ref{fig.Scans}. 

\bigskip
{\em Calorimetry}

Differential Scanning Calorimetry 
studies have been performed with a DNA film identical to
the one used for neutron scattering, but prepared from a different
DNA solution. Two different samples were used, with masses of $49\;$mg
and $45.5\;$mg. The samples cut in the film 
were rolled and hermetically sealed into the
hastelloy sample tube of a Setaram Micro DSC III calorimeter. The
reference tube of this differential calorimeter was empty. After a
cooling to $278\;$K the temperature $T$ has been raised to
$368\;$K or $383\;$K at a rate of $0.6\;$K/min, 
maintained for $10\;$min at the maximum
temperature and decreased to  $278\;$K at the same rate of
$0.6\;$K/min. 
The differential heat flux $\Delta\phi$ has been measured as a
function of time (temperature) and the specific heat has been obtained
from $\left(\Delta\phi+\tau\frac{d\Delta\phi}{dt}\right)/(dT/dt)$
where $\tau$ is the thermal time constant of the calorimeter (here
$\tau=60s$) \cite{Hohne}. 

\bigskip
{\em Optical observations}

A small piece of DNA film has been sealed between two glass plates to
preserve its water content while it was heated on a hot plate 
below an optical microscope at the rate of $1\;$K/min. The
sample was lighted and observed from the top as it went through the
denaturation temperature of DNA.

\bigskip
{\em Gel electrophoresis}

A small piece of the sample (0.01~g taken
before and after heating in the neutron scattering experiment) 
was dissolved in water. The solution has be used to run a standard gel
electrophoresis experiment, using a 1\% non-denaturing
agarose gel and stained with Ethidium Bromide.
Comparisons with DNA mass ladders were used to 
measure the length of the DNA fragments in the sample. 

A similar
experiment was performed with the solution used to prepare the DNA
fibres to probe the state of the DNA molecules prior to any
treatment.

\subsection{Experimental results}
\label{subsec:expresults}

\begin{figure}
\begin{tabular}{c}
\includegraphics[width=7cm]{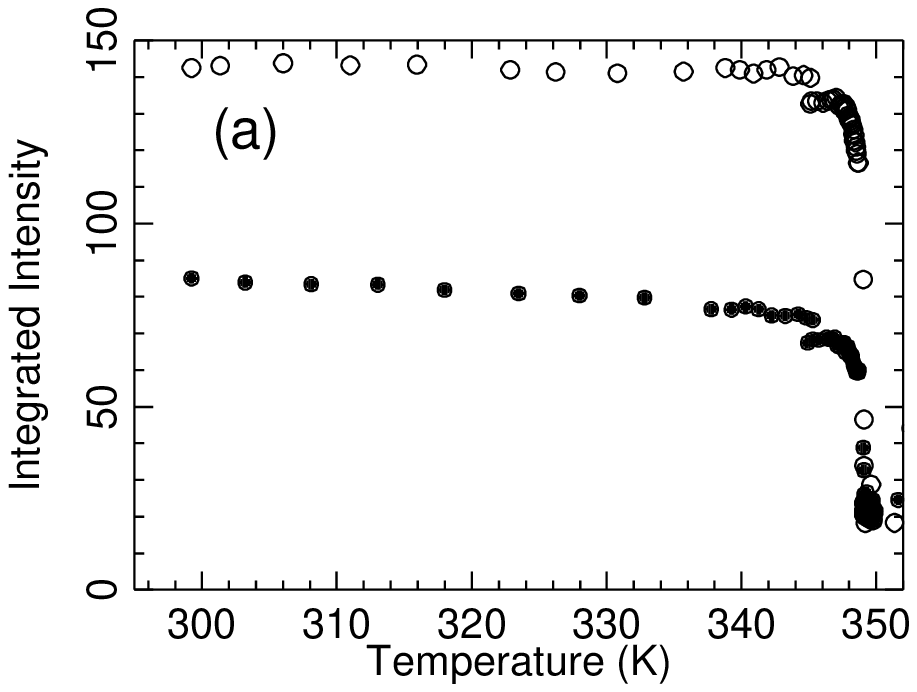} \\
\includegraphics[width=7cm]{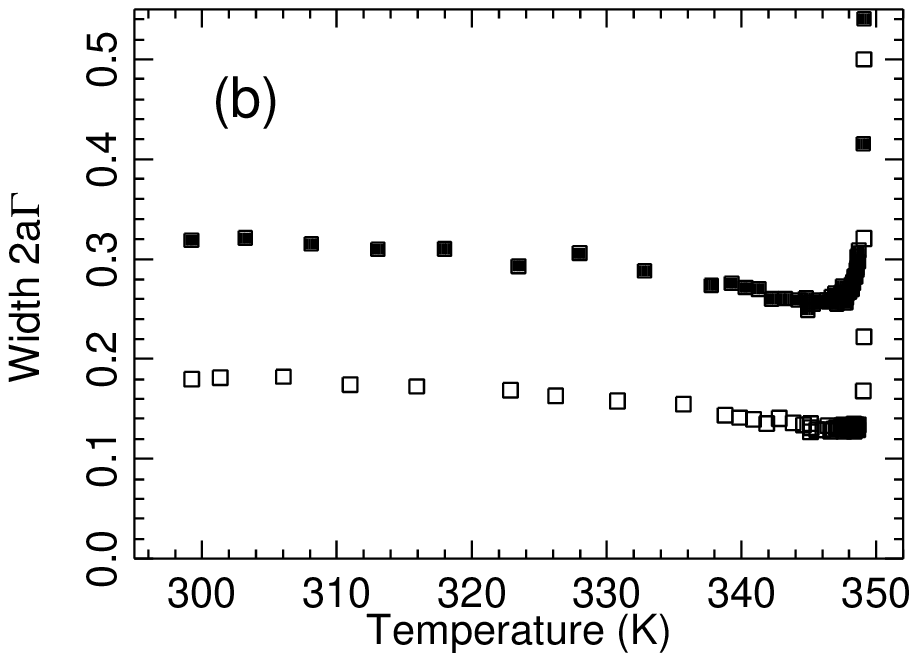} \\
\includegraphics[width=7cm]{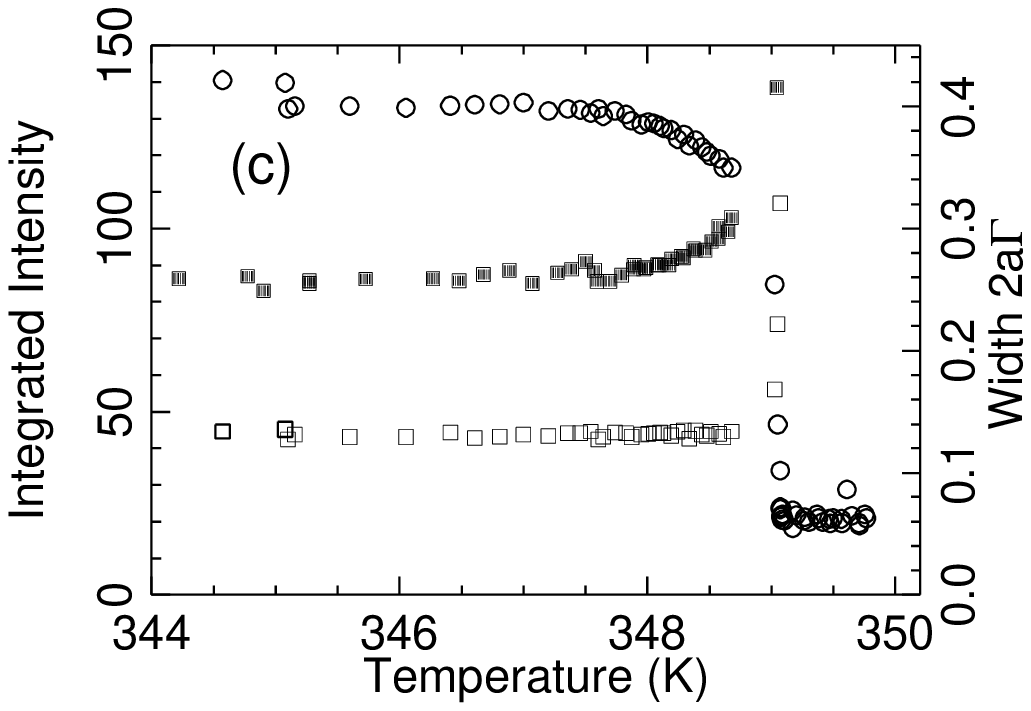} 
\end{tabular}
\bigskip
\caption{
Integrated intensity (a) and width $2 a \Gamma$ (dimensionless) (b)
of the Bragg peaks versus temperature.  
 The small discontinuity at $T\sim345\;$K is due to a small
 misalignment of the instrument that was discovered and subsequently
 corrected.  Results for scan 1 use empty symbols 
while results for scan 2 are plotted with filled symbols. 
The bottom panel (c)
shows the temperature evolution of the intensity (circles)
and width of the peak in the immediate vicinity of the melting
transition. The evolution of the intensities
are the same for both scans, hence data is only shown for scan 1.  The
widths are shown for scan 1 (open squares) and scan 2 (closed
squares).  }
\label{fig:ampliwidth}
\end{figure}

\begin{figure}
  \includegraphics[width=7cm]{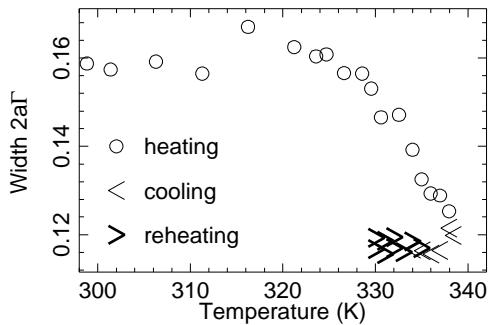}
  \caption{Experiment showing the annealing of the sample due to
    heating to moderate temperature. The data correspond
    to a scan of type 1. The Bragg peak gets sharper on
    heating (open circles) and keeps its lower width if it is
    subsequently cooled and heated again.}
\label{fig:heat-cool} 
\end{figure}

Figures  \ref{fig.Scans} and \ref{fig:ampliwidth} illustrate the main
result given by neutron diffraction. The integrated intensity of the
Bragg peak stays constant from room temperature to about $T =
346\;$K. At this temperature it starts to show a small decrease
occurring on a temperature range of about $3\;$K, followed by an abrupt
drop. A more careful examination of Fig.~\ref{fig:ampliwidth} exhibits
the following results:
\begin{itemize}
\item The intensity of the peak observed in scan 1 is larger than that
  of the off-centre peak in scan 2, as expected, but its evolution
  versus temperature is remarkably similar in both scans.
\item The width of the off-centre peak (scan 2) is significantly
  larger than the width at the centre of the diffraction spot.
\item In the $300 - 340\;$K temperature range the width of both peaks
  is essentially constant and even shows a slight decrease which can
  be attributed to an annealing of the sample as shown by
  Fig.~\ref{fig:heat-cool}. In the experiment shown in this figure,
  another sample was heated up to $340\;$K then cooled down to
  $330\;$K and heated up to $340\;$K again. During the first heating
  the width of the peak decreases, but then, in the cooling stage it
  keeps its lowest value, as well as during reheating. In this
  experiment the integrated intensity of the peak stays constant for all
  temperatures.

\item For scan 1, Fig.~\ref{fig:ampliwidth}-c 
  shows that, in the vicinity of the transition, where the annealing
  has been completed, the width of
  the Bragg peak is remarkably constant until  the temperature where
  the intensity drops 
  abruptly. This width does not show any precursor effect, even when
  the intensity of the mode starts to decrease in the temperature
  range $346 < T < 349\;$K. On the contrary for scan 2 (off-centre)
  the width of the peak shows a gradual increase in this temperature
  range, which appears to mirror the decrease of the intensity of the
  peak.
\end{itemize}

\begin{figure}
  \begin{tabular}{c}
\includegraphics[width=7cm]{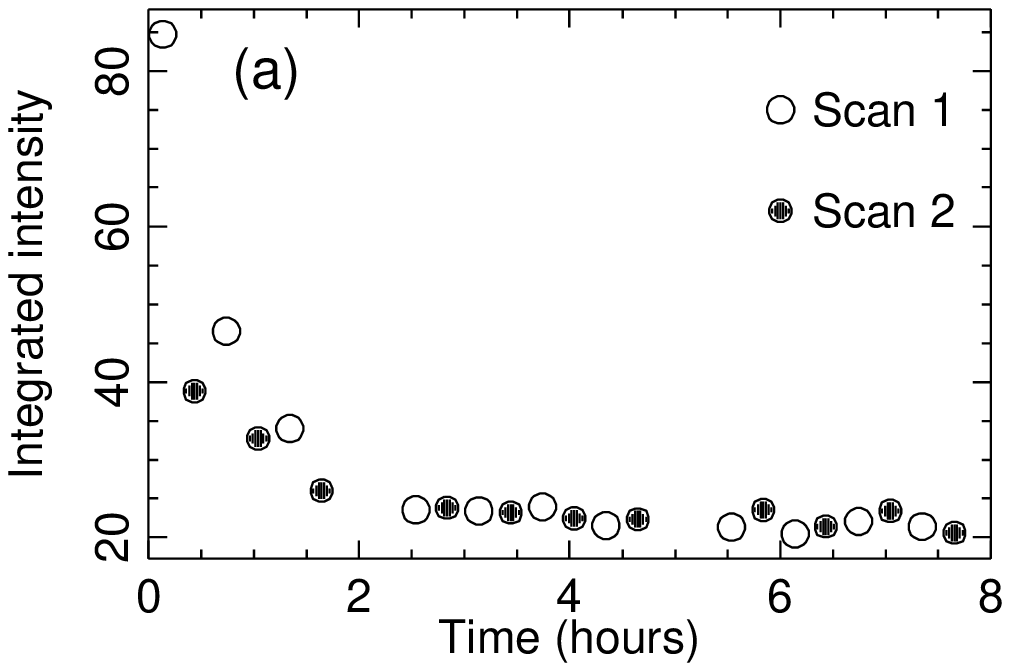} \\
\includegraphics[width=7cm]{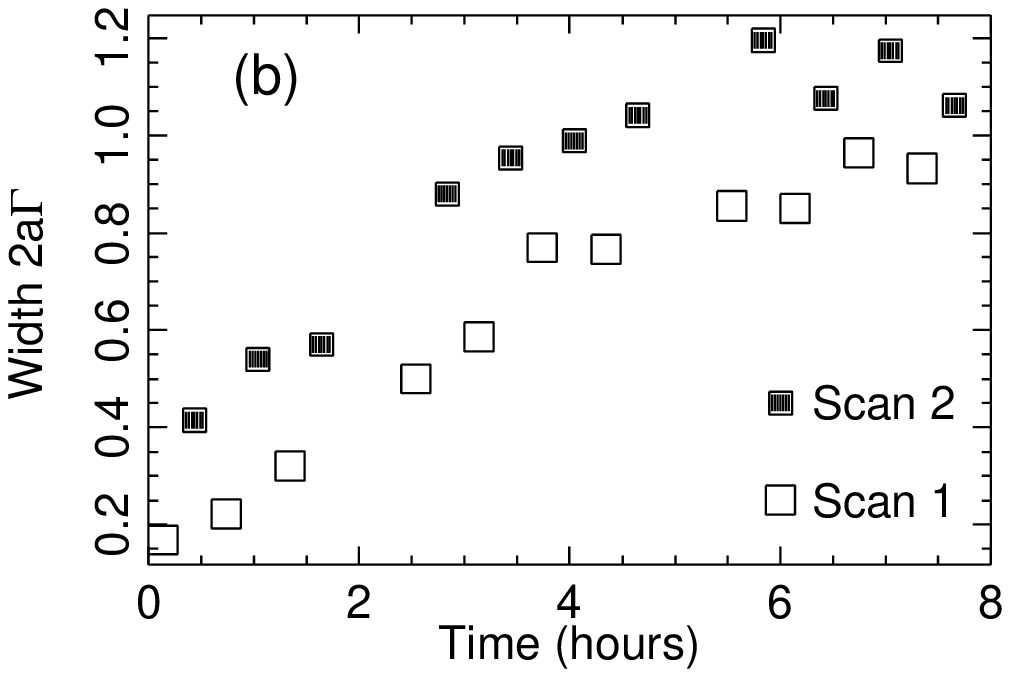} 
\end{tabular}
  \caption{Time dependence of the integrated intensity (a) and width
    (b) of the Bragg peak at fixed temperature $T=349.05 \pm 0.03\;$K
    where the sharp drop of the peak intensity occurs.}
  \label{fig:timedep}
\end{figure}

The drop of the Bragg-peak intensity, shown in
Fig.~\ref{fig:ampliwidth} is extremely sharp in terms of temperature,
but requires some time to complete. This is shown in
Fig.~\ref{fig:timedep}. When the transition was reached and the intensity
started to drop, we stopped cooling and observed the time evolution of
the peak at constant temperature $T=349.05 \pm 0.03\;$K. 
Figure~\ref{fig:timedep} shows that the intensity needed about 3 hours
to stabilise to a low value, and the width was not stabilised before
about 6 hours.

\begin{figure}
  \includegraphics[width=8cm]{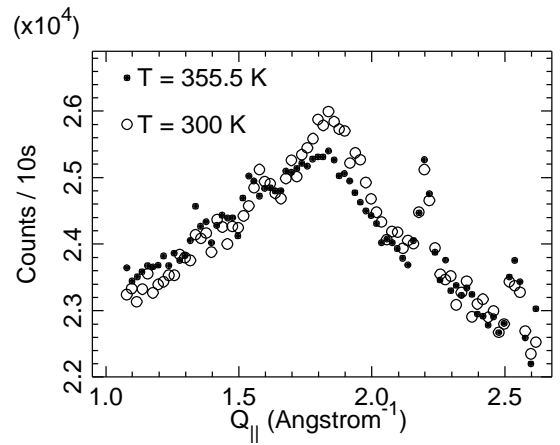}
  \caption{Plot of the Bragg peak remaining after heating the sample
    to $355\;$K (closed symbols) and after subsequent cooling to room
    temperature (open symbols). The sharp peak around
    $2.2\;$\AA$^{-1}$ is due to the aluminium of the sample holder.}
  \label{fig:reversibility}
\end{figure}

In the fibre sample the
transition that corresponds to the almost complete vanishing of
the Bragg peak associated to the stacking of the base pairs is not
reversible. On cooling we did not observe a reappearance of the peak,
although a very small recovery can be observed on a magnified picture
of the Bragg peak recordered at room temperature after cooling as
shown in Fig.~\ref{fig:reversibility}.

\smallskip

\begin{figure}[h]
\includegraphics[width=9cm]{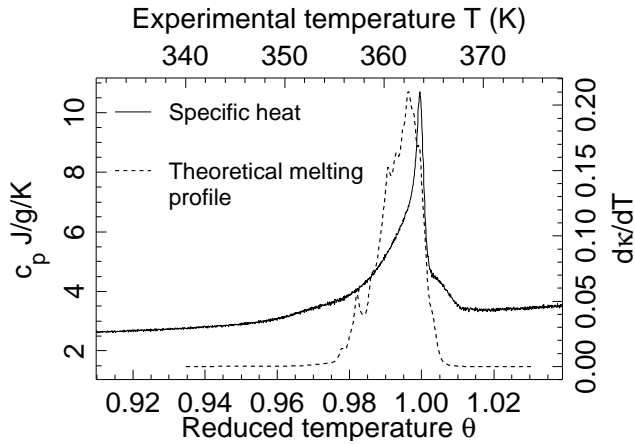}
  \caption{Specific heat, $c_p$, of a B-form Li-DNA film
    similar to the film used in neutron scattering experiments (full
    line, left scale) obtained by DSC. The dash line shows the theoretical
    denaturation profile of the sequence used for the analysis 
    (Section \ref{sec:theory}) :
    derivative with respect to temperature of the
    fraction $\kappa$ of open base pairs.
The figure is identical to Fig.~1 \cite{PRL}.}
  \label{fig:calo}
\end{figure}

Figure \ref{fig:calo} shows the specific heat of a DNA film obtained
by Differential Scanning Calorimetry (DSC). The sharp peak at
$360\;$K can be attributed to the thermal denaturation of DNA although
the transition temperature $T_c$ cannot be quantitatively compared with
neutron observations because the measurement was done on a different
sample. $T_c$ strongly depends on
external conditions, and particularly the ionicity of the solvent
\cite{Kamenetskii} so that the shape of the melting curve is
more significant than the value of the denaturation temperature
when comparing samples.
The transition is not reversible, and the sharp peak does not reappear if
the sample is cooled and the measurement repeated \cite{DNAglass}.

\begin{figure}
   \begin{tabular}{cc}
     \textbf{(a)} &  \textbf{(b)} \\
\includegraphics[width=4cm]{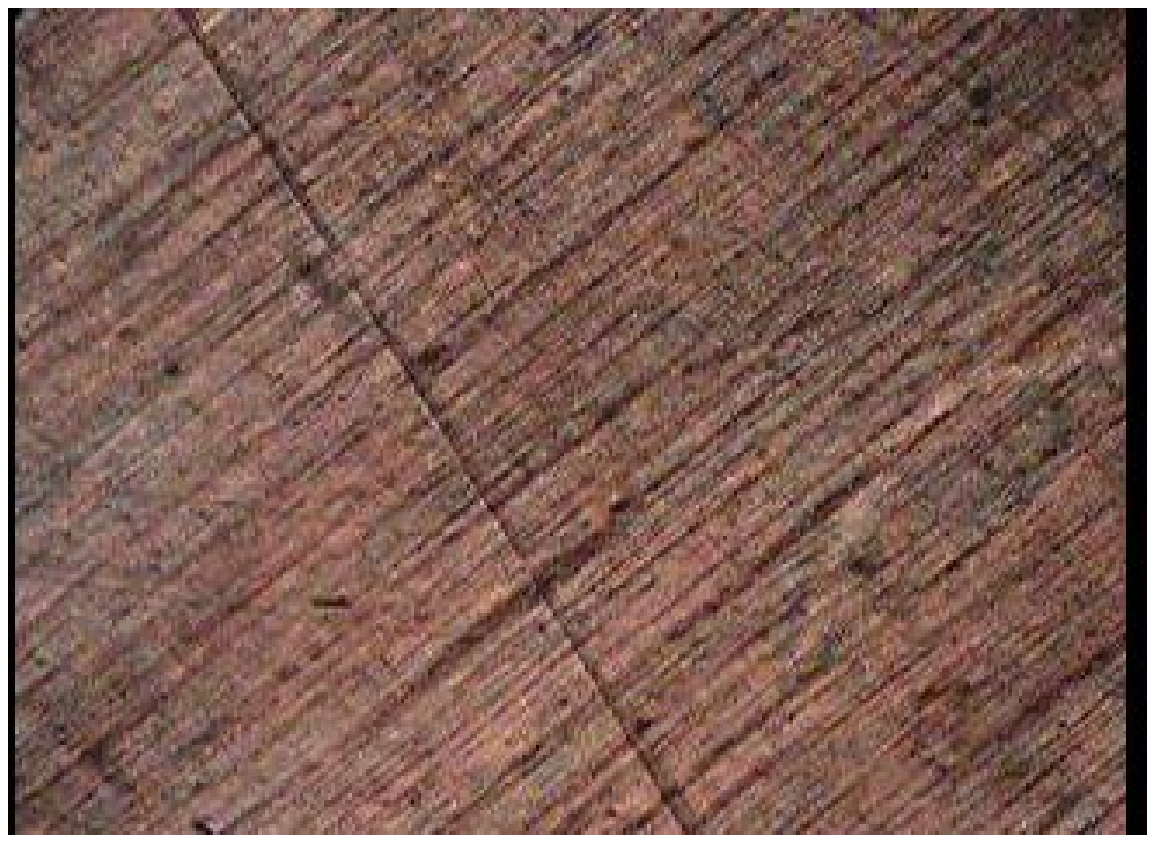} &
\includegraphics[width=4cm]{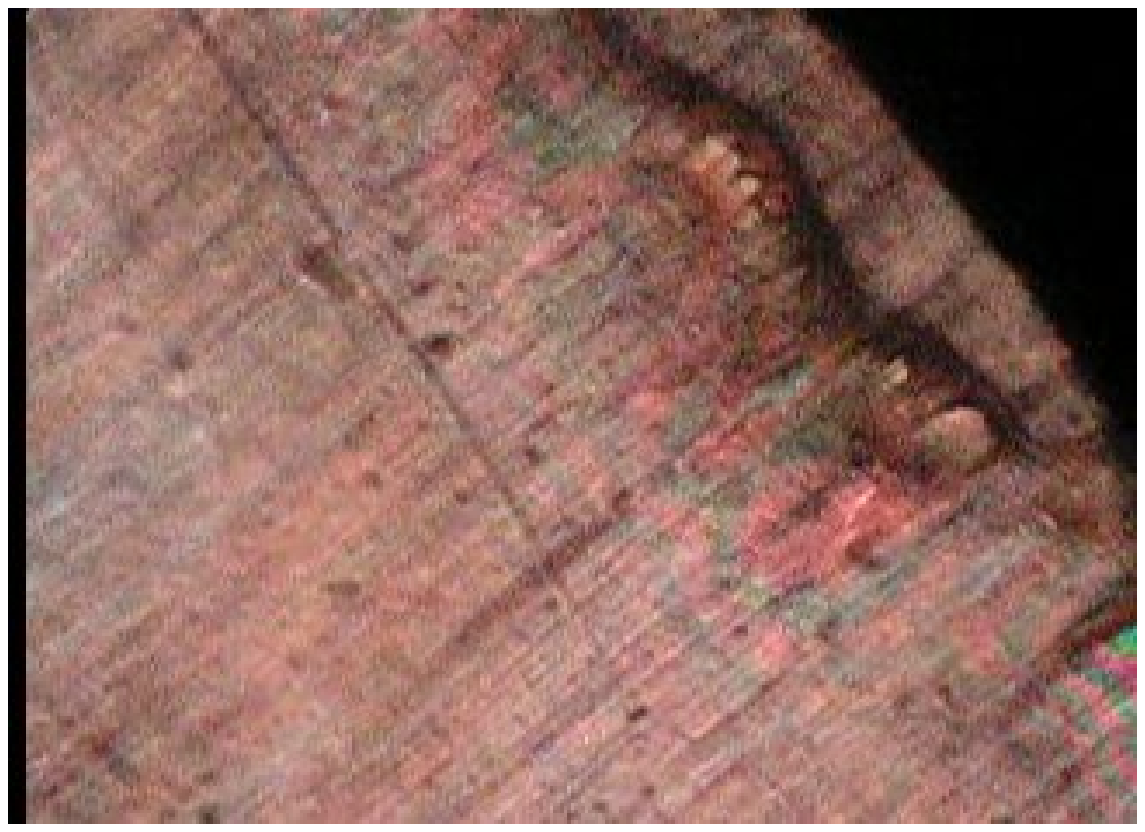} 
   \end{tabular}
   \caption{Optical microscopy observation of a piece of film prior (a)
     and after (b) heating above the denaturing temperature of
     DNA. }
\label{fig:optical}
 \end{figure}

\smallskip
 The fibre structure of the film is clearly visible in optical
 microscopy observations of heated DNA films until the denaturation
 temperature is reached. Then this organised structure of parallel
 fibres is essentially lost and the film tends to shrink
 (Fig.~\ref{fig:optical}).

\begin{figure}
  \includegraphics[width=6cm]{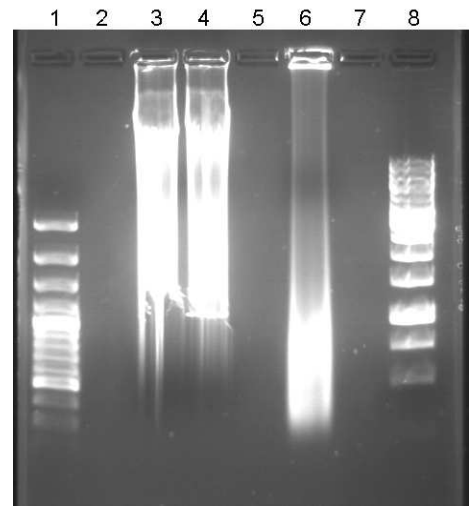}
  \caption{Electrophoresis image showing the length of the DNA molecules
    before and after heating of the film. From left to right:
DNA mass ladder SM0321 (100-3000 base pairs) (lane 1), 
solution used to prepare
the DNA film (lane 3), 
solution prepared from a piece of film that has not been
heated (lane 4), solution prepared from a piece of film after
heating (lane 6),
DNA mass ladder SM0311 (250-10000 base pairs) (lane 8). Lanes
2,5,7 were not used.}
\label{fig:electrophoresis}
\end{figure}

\smallskip
Gel electrophoresis 
shows that the length of the DNA molecules in the solution used to
prepare the film, or in a piece of film which has not been heated, is
of the order of 20 kilo-base or larger. However the same measurements
performed on a piece of film which has been heated up to DNA
denaturation and film collapse, and subsequently cooled to room
temperature, only detects DNA fragments of a few hundreds of bases,
indicating that the original molecules have been chopped in the
thermal cycle (Fig.~\ref{fig:electrophoresis}).

\section{Analysis}
\label{sec:theory}

To analyse the neutron diffraction results, we need to proceed in two
steps. First, we must determine the diffraction pattern of a finite 
segment
of double stranded DNA, taking into account the local inhomogeneities
in its structure which are associated to base pair sequence, and the
thermal fluctuations. Second we must study the statistical physics of
DNA to determine the size distribution of the closed segments
of DNA as a function of temperature.

\subsection{Structure factor of a closed DNA segment.}

We consider a structurally disordered linear chain of $M$ sites. Let
$a$, the average base-pair spacing, 
be the average distance between successive sites and $\lambda_j$
the local structural deviation from that value between $j\,$th and
$(j+1)\,$st sites; the equilibrium positions of $m\,$th and $n\,$th sites
differ by $(m-n)a +\sum_{j=n+1}^m \lambda_m$. Structural disorder in
the transverse direction is similarly expressed by $\eta_j$, the local
deviation (from zero) of the distance between $j\,$th and
$(j+1)\,$st sites in the transverse direction.
The structure factor of such a finite chain segment is given by  
\begin{align}
\label{eq:defsq}
S_M(\mathbf{Q})=\frac{1}{M}\sum_{m,n=1}^{M} & e^{iQ_{\parallel} a (m-n)}
\langle e^{  i \sum_{j=m+1}^n (Q_{\parallel} \lambda_j + Q_{\perp} \eta_j) }\rangle \nonumber \\
& \times \langle e^{iQ_{\parallel} (x_m - x_n) +iQ_{\perp} (y_m-y_n)} \rangle
\end{align}
where $x_m, y_m$ represent, respectively,
the longitudinal and transverse displacements of the $m\,$th site from
its position at thermal equilibrium, and $\mathbf{Q}$ the scattering
vector, having the component $Q_{\parallel}$ along the helix axis and $Q_{\perp}$
orthogonal to it.
Equation (\ref{eq:defsq}) is a slight generalisation of the finite
para-crystal theory \cite{parax} to account for disorder and motion in
the transverse direction. The angular brackets denote averages which
can be decoupled because the first refers to structural disorder and
the second to thermal motion.
 To a first approximation, structural
disorder is modelled by Gaussian variables
$\{\lambda_j\}$ and $\{\eta_j\}$ with zero average and
\begin{eqnarray}
\label{eq:Gaussvar}
\langle\lambda_i \lambda_j\rangle & = & 
\langle\lambda^2\rangle \> \delta_{ij} \\ \nonumber
\langle\eta_i \eta_j\rangle & = & 
\langle\eta^2\rangle \>  \delta_{ij}  \\ \nonumber
\langle\eta_i \lambda_j\rangle &=& 
\chi [\langle\eta^2\rangle\langle\lambda^2\rangle]^{1/2} \> \delta_{ij} 
\end{eqnarray}
where $\delta_{ij}$ is the Kronecker symbol ($\delta_{ij} = 1$ if
$i = j$ and $0$ otherwise).
We use estimates of the variances $\langle\lambda^2\rangle$ 
and $\langle\eta^ 2\rangle$ obtained from conformational analysis \cite{Lavery}
and present alternative calculations for uncorrelated ($\chi=0$) and
correlated ($\chi \not= 0$) structural disorder. Thermal fluctuations in 
the longitudinal displacements can be calculated in the harmonic
approximation; 
thermal fluctuations in the transverse direction can also be 
calculated  (cf. next subsection for a particular model). We will
describe them in an approximate fashion which 
takes into account the thermal effects due to sequence heterogeneity,
i.e. local variations in $\xi_j=\langle y_{j+1}\rangle-
\langle y_j\rangle$; then  
\begin{equation}
\label{eq:transav}
\langle e^{iQ_{\perp} (y_m-y_n)}\rangle \approx e^{-Q_{\perp}^ 2\langle\xi^2 \rangle|m-n|/2}  
\end{equation}
where $\langle\xi^2\rangle$ is the variance of the $\{\xi_j\}$'s.

Putting the various terms together and performing one of the two
summations \cite{parax} results in 
\begin{align}
\label{eq:sn}
  S_M({\mathbf{Q} }) &= M + 2 \sum_{n=1}^{M-1} (M-n) \cos (Q_{\parallel} na) 
e^{- n \Delta}
 \; ,
\end{align}
where 
\begin{align}
2\Delta = Q_{\parallel}^2 (\langle\lambda^2
\rangle+ \langle \sigma^2 \rangle)&+
Q_{\perp}^2(\langle\eta^2\rangle+\langle\xi^2\rangle) \nonumber \\
&+ 2 Q_{\parallel} Q_{\perp} \chi (\langle\eta^2\rangle
     \langle\lambda^2\rangle)^{1/2} 
\end{align}
and 
$\sigma^2=k_B T/\mu c_0^2 a^2$,
the Debye-Waller correction due to longitudinal thermal motion
at any temperature $T$, can be obtained from the total DNA
mass per base pair $\mu = 618\;$a.m.u. and the measured 
\cite{DNABrill} sound
velocity $c_0 = 2830\;$m.s$^{-1}$;  $k_B$ is the Boltzmann constant.  

Near the first Bragg peak, which is where the present experiment
focused, and for sufficiently large cluster sizes, the sum
(\ref{eq:sn}) 
can be approximated by
\begin{equation}
\label{eq:sn1bragg}
  S_M^*({\mathbf{Q}}) \sim M {\cal S}'(\mathbf{Q}) \doteq M \frac{\sinh\Delta} 
{\cosh\Delta - \cos(Q_{\parallel} a)}
  \; ,
\end{equation}

\subsection{Statistical physics of the closed regions of DNA.}

To calculate the size distribution of the closed segments of DNA by a
statistical physics analysis, we selected the PBD model \cite{PBD}
which is sufficiently simple to allow the analysis of DNA segments of
thousands of base pairs, but nevertheless includes some data on the
structure of the molecule which are necessary to calculate the
structure factor. Moreover, as it describes the molecule in terms of a
Hamiltonian, its parameters are directly linked to physical
quantities. Therefore they are easier to determine than for Ising
models, although they still need to be refined by comparison with a
variety of experimental melting profiles.

\medskip
The configuration energy of a DNA molecule of $N$ base pairs is
written as
\begin{equation}
  \label{eq:HPBD}
H_y = \sum_{j=1}^{N-1} W(y_j,y_{j+1}) + \sum_{j=1}^{N} V_j(y_j)
\end{equation}
where $y_j$ represents the stretching of the $j^{th}$ base pair, due to
the transverse displacements of the bases. The
stacking interaction between adjacent bases is described by the
anharmonic potential \cite{PBD}
\begin{equation}
  \label{eq:W}
W(y_j,y_{j+1}) = \dfrac{1}{2} k \left[ 1 + \rho e^{-b(y_j +
    y_{j+1})}\right](y_j - y_{j+1})^2
\end{equation}
which takes into account the weakening of the interactions when the
pairs are broken. The potential $V_j(y_j) = D_j[1 - \exp(- \alpha_j
  y_j)]$ is a Morse potential which describes the combined effects of
hydrogen-bonding, electrostatic interactions between the charged
phosphate groups, and solvent effects on the the $j^{th}$ base pair.
The 4 possible bases, $A$, $T$, $G$, $C$ form two types of pairs,
$A-T$, linked by two hydrogen bonds, and $G-C$, linked by three
hydrogen bonds. Both the stacking interactions $W$ and the intra-pair
potential $V$ are affected by the sequence of bases. However,
although subtle sequence effects on short DNA fragments may require
the introduction of different stacking potentials for different
interacting pairs \cite{JPCM}, the melting curves of long DNA molecules,
with several thousands base pairs, can be accurately reproduced by
introducing the effect of the sequence in the intra-pair
potential $V_j$  only \cite{NTh}, which drastically reduces 
the number of model
parameters. Therefore in our calculations the stacking interaction is
treated as homogeneous.

\medskip
To determine the melting curve or calculate the size of the closed
regions, we need to give a quantitative definition of a closed base
pair. This can be done by choosing a reference stretching $y_c$. Base
pair $j$ is considered as closed if $y_j < y_c$. We select $y_c =
1.5\;$\AA, which corresponds to a base pair whose stretching is well on the
plateau of the Morse potential. The results are weakly dependent of
the value of $y_c$ provided it is larger than $1\;$\AA\ because
molecular dynamics simulations show that, once
a base pair as been stretched to a value that brings it on the plateau
of the Morse potential, it is likely to open widely.

\medskip
The statistical weight of a given configuration of the molecule
thermalised at temperature $T$ is
\begin{align}
  \label{eq:z1}
{\cal Z} = \int_{y_{l_1}}^{y_{m_1}} dy_1 \int_{y_{l_2}}^{y_{m_2}} dy_2 &\ldots
\int_{y_{l_N}}^{y_{m_N}} dy_N \nonumber \\
& e^{- H_y(y_1, y_2, \ldots y_N) / (k_B T)}   \; . 
\end{align}
The
limits of the lower and upper bounds of the integrals $y_{l_j}$,
$y_{m_j}$ depend on the particular configuration. Setting $y_{l_j} =
-\infty $, $y_{m_j} = + \infty $ for all $j$ allows the molecule to
explore its full configurational space and ${\cal Z}$ is then the
partition function $Z$. Setting $y_{l_j} =-\infty $, $y_{m_j} = y_c$
defines a configuration in which base pair $j$ is closed while
 $y_{l_j} = y_c $, $y_{m_j} = + \infty$ defines a configuration where
it is open. The integrals associated to all those configurations can
be easily calculated because the model is one-dimensional and
restricted to nearest-neighbour coupling so that, instead of a highly
multidimensional integral, one has to compute a chain of
one-dimensional integrals \cite{NTh,vanerp,zhang} involving a kernel
which depends on the stretching at two adjacent sites. Moreover the
speed of the calculation can be significantly increased by expanding the
site-dependent kernel on the basis of the eigenfunctions of a
reference kernel, for instance  the kernel associated to an $A-T$ base
pair \cite{NTh}.

These calculations allow us to obtain the probability ${\cal P}(m,i)$
that $m$
adjacent sites, starting at site $i$ are closed,
\begin{align}
  {\cal P}(m,i) = \dfrac{1}{Z} {\cal Z}(y_1, y_2, & \ldots y_{i-1}, y_i < y_c,
y_{i+1}<y_c, \ldots, \nonumber \\
& \ldots y_{i+m-1} <y_c, y_{i+m}, \ldots ) \; ,
\end{align}
by computing the statistical weight of configurations
where restrictions on the integration range are imposed for all sites
belonging to that closed region and no restrictions are imposed
elsewhere. Then these quantities give the probability $P(m,i)$
that a closed cluster of size $m$, with open ends, exists at site $i$
through 
\begin{align}
  P(m,i) = {\cal P}(m,i) &- {\cal P}(m+1,i-1) \nonumber \\ 
&- {\cal P}(m+1,i) + {\cal P}(m+2, i-1)
\end{align}
from which the probability to have a closed cluster of size
$m$ in a DNA segment of $N_0$ base pairs is simply
\begin{equation}
  P(m) = \dfrac{1}{N_0} \sum_{i=1}^{N_0} P(m,i) \; .
\end{equation}
The average size of a cluster of closed base pairs is then
\begin{align}
\label{eq:lc}
  l_c = \dfrac{\sum m P(m)}{ \sum P(m)} =
\dfrac{h}{1 - h - P_0} \; ,
\end{align}
where $h$ is the helix fraction and $P_0$ the
statistical weight of two consecutive bases being in the closed state.
Similar calculations for the open regions of the
DNA molecule give the average size of the denatured regions $l_o = (1 - h)/
(1 - h - P_0)$.
These quantities are computed for a range of cluster sizes from $1$ to
a maximum value $M$.
To avoid end effects we study a DNA segment of size $N = N_0 + 2 M$
and the sites $i$ are chosen so that the clusters that we consider
are formed of the bases pairs which are at least $M$ sites away from
the ends.

\bigskip
For a natural DNA sample, which may contain millions of base pairs,
the intensity observed in a neutron scattering experiment is
proportional to
\begin{equation}
  \label{eq:sq}
S({\mathbf{Q}}) = \sum_{m=1}^{\infty} P(m) S_m({\mathbf{Q}})
\end{equation}
where $S_m({\mathbf{Q}})$ is given by Eq.~(\ref{eq:sn}), and the size
of the DNA molecule has been extrapolated to infinity.

In practice the calculation is performed with segments of natural DNA
which include $280 000$ base pairs, and $P(m,i)$ is computed up to
cluster sizes $M$ of a few hundreds of base pairs (typically
$M=150$). It is however easy to determine
$P(m)$ for large $m$ because it scales exponentially with $m$
\cite{NikosPRE}, as shown in Fig.~\ref{fig:theomelt}-b. Fitting the
numerical data of $\log(P(m)$ for $40 \le m \le M$ by a straight
line we get $P(m) = P_0 \zeta^m $ for $m>M$ so that in practice the
calculation of $S({\mathbf Q})$ by Eq.~(\ref{eq:sq}) is expressed as
\begin{align}
  \label{eq:sq2}
S({\mathbf{Q}}) =  \sum_{m=1}^{M} & P(m) S_m({\mathbf{Q}}) + 
\sum_{m=M}^{M_0} P_0 \zeta^m
S_m({\mathbf{Q}}) \nonumber \\ 
&+ {\cal S}'({\mathbf{Q}}) P_0 \zeta^{M_0 + 1} \left[ \dfrac{M_0}{1 - \zeta} +
  \dfrac{1}{(1 - \zeta)^2} \right] \;,
\end{align}
where the summation for $m > M_0$ ($M_0 = 1000$) 
has been calculated analytically
using the property that, for large $m$, 
$S_m({\mathbf{Q}})$ can be approximated by the limiting form
(\ref{eq:sn1bragg}). 
The same method can be used to compute the average cluster size
$l_c$ (Eq.~(\ref{eq:lc})) versus temperature. A typical result is shown on
Fig.~\ref{fig:theomelt}-c. Then the structure factor is fitted with
the same Lorentzian expression as the one used to analyse experimental
data to determine the integrated intensity
$I_0$ and width $\Gamma$ of the diffraction peak.

\subsection{Model parameters}

To analyse of the neutron scattering experiments, in principle 
we would need to know the base-pair sequence in the sample. 
As the experiment requires a significant amount of DNA
it can only be performed with natural
DNA. The salmon testes DNA that we use is provided without its
sequence \cite{Datasheet}, and even its $GC$ content is only
approximately known. It is estimated to be 41.2\%. 

The theoretical analysis has been tested on different DNA sequences
from the genome of {\it Danio rerio} (zebrafish) \cite{Genebank} 
and {\it Pyrococcus abyssi} \cite{Pabyssi}. The results presented in
Fig.~\ref{fig:theomelt} have been obtained with a sequence of 
280000 bases, part of the full genome of 
{\it Pyrococcus abyssi}, chosen because
its denaturation curve is the closest to the denaturation profile
of our samples measured by differential scanning calorimetry
(Fig.~\ref{fig:calo}). The $GC$ content of
this fragment is 44.08\%

Model parameters have been obtained from an extensive study of DNA
denaturation on various sequences \cite{NTh}, which determined a set
of parameters allowing the prediction of melting curves of
various DNA sequences to a good accuracy. Experiments to
record DNA denaturation curves are generally performed in solution
with sodium salt, the Na$^+$ ions being necessary to stabilise DNA by
screened the repulsions between the charged phosphate groups.
However our neutron scattering
experiments have been performed with Li-DNA because its secondary structure
is more stable, and stays in the B-form over a much higher humidity range,
than that of Na-DNA \cite{Korolev,Lindsay}. 
Therefore the parameters of the $AT$ and $GC$ Morse
potentials cannot be determined unambiguously. The values that we use
correspond to a high sodium content. They have been chosen on the basis
of the shape of the melting curve rather than the value of the
denaturation temperature $T_c$. This is why the comparison between theory
and experiments must be made with the reduced temperature
$\theta = T / T_c$ rather than with the actual temperature.
The parameters selected for the analysis are $k = 0.00045\;$eV/\AA$^2$,
$\rho = 50$ and $b = 0.20\;$\AA$^{-1}$, 
$\alpha_{AT} = 4.2\;$\AA$^{-1}$, $\alpha_{GC} = 6.9\;$\AA$^{-1}$
and $D_{AT} = 0.12905\;$eV, $D_{GC} = 0.16805\;$eV.

\subsection{Results}

\begin{figure*}[h]
\begin{tabular}{cc}
\textbf{(a)} & \textbf{(b)}\\
\includegraphics[height=6cm]{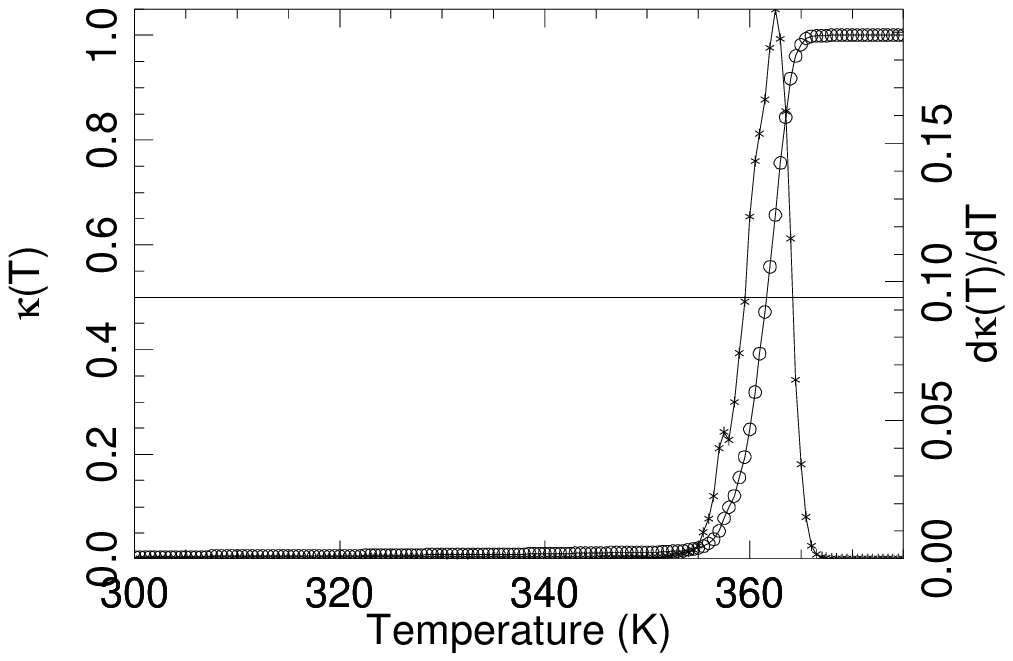} &
\includegraphics[height=6cm]{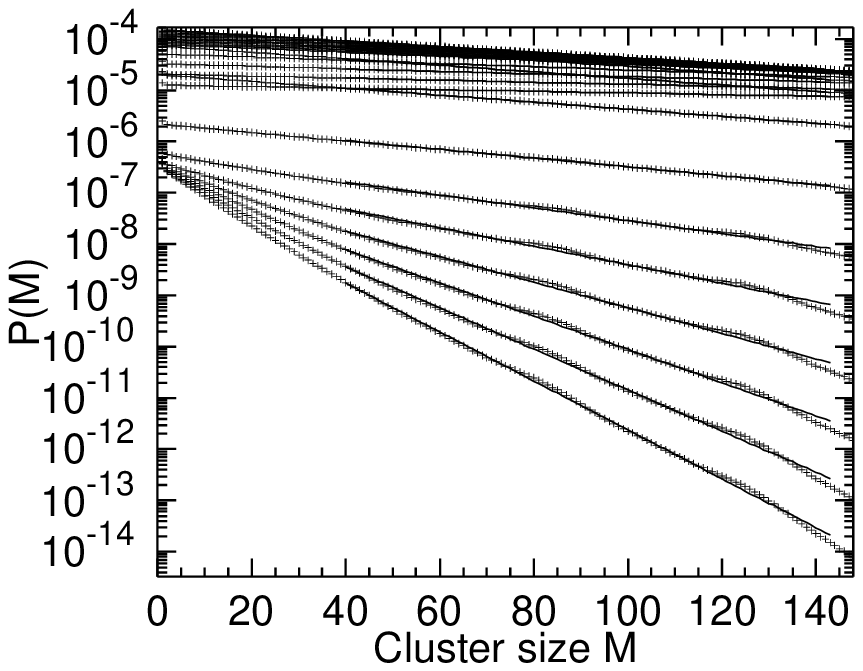} \\
\textbf{(c)} & \textbf{(d)}\\
\includegraphics[height=6cm]{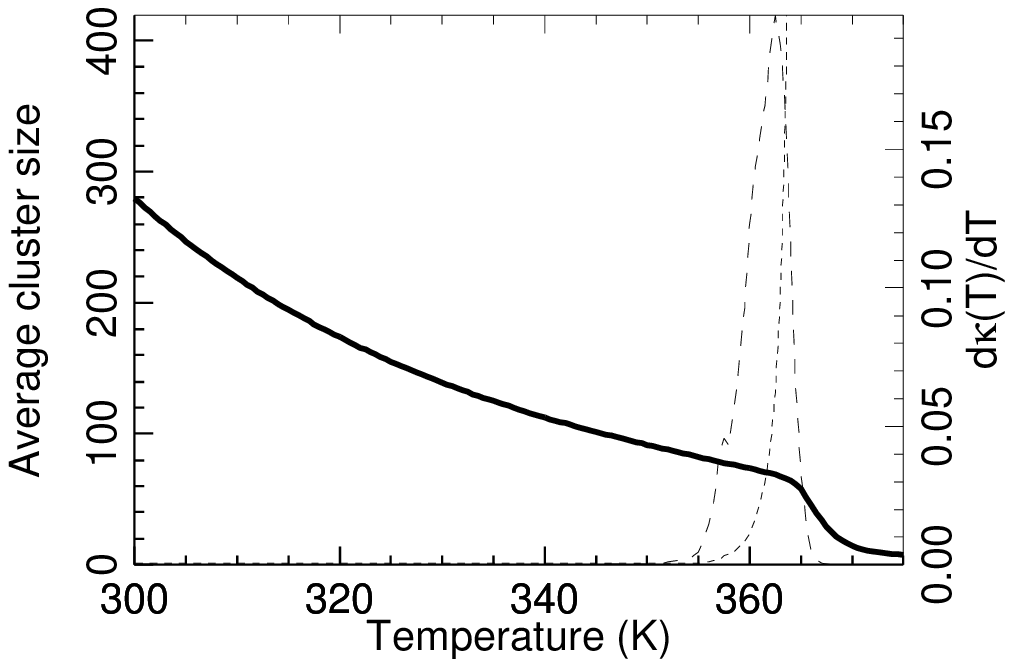} &
\includegraphics[height=6cm]{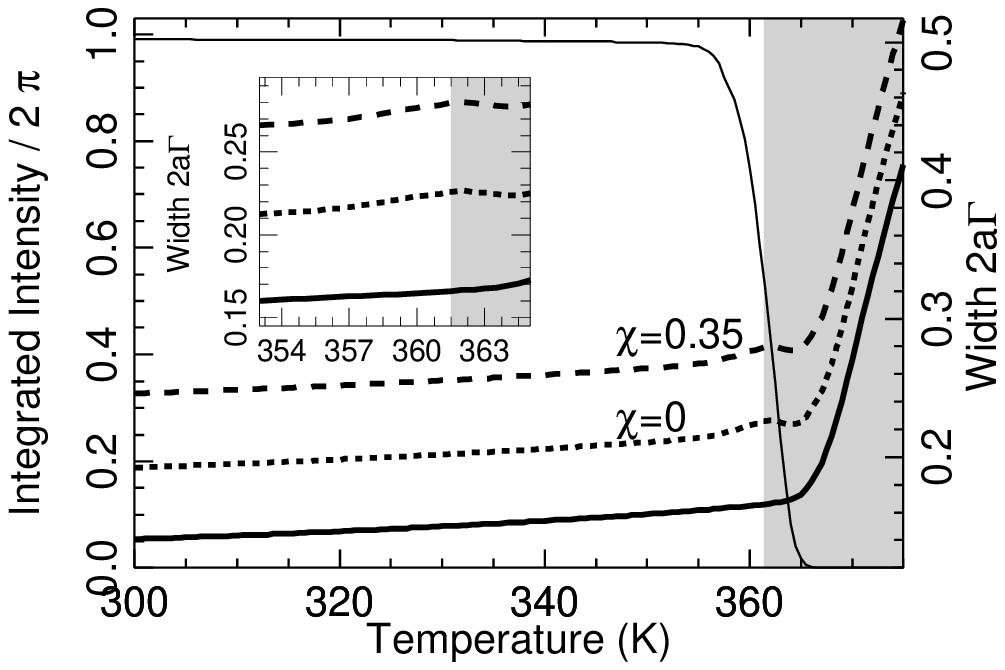} 
\end{tabular}
\bigskip
  \caption{ Theoretical results.
(a) Melting curve of a reference DNA segment of 280000 base
  pairs, part of the genome of {\it Pyrococcus abyssi}: the circles show the
open fraction versus temperature and the stars correspond to its
derivative to get the melting profile.
(b) Probabilities $P(m)$ versus $m$ at different temperatures in the
  range $T=330\;$K to $T=375\;$K, in logarithmic scale. The points are
  the values calculated from the statistical mechanics of the DNA
  sample, and the lines show a linear fit for $40 \le m \le 150$. 
  Curves are plotted every $10\;$K from 300 to $340\;$K and every
$1.5\;$K above. (c)
  The average size the closed clusters $l_c$ (full line) and the average
size of the open regions $l_o$ (dotted line) versus $T$. The dashed line
shows the melting profile on the same temperature scale.
(d) Integrated intensity (thin full line) and width (thick full line
for scan 1,
  dotted and dashed lines for scan 2) versus temperature. For scan 2 the
  figure shows two cases: $\chi = 0$, i.e. ignoring the correlation
  between longitudinal $\lambda_j$ and transverse $\eta_j$ 
components of the structural  disorder due to the sequence (dotted
line) and $\chi = 0.35$ (dashed line) 
which assumes a partial correlation. 
The shaded area of the plot shows the temperature range in which the
experimental observations are hindered by the melting of the sample
film (Sec.~\ref{sec:discussion}). The inset shows a magnification of
the variation of the widths versus temperature in the immediate
vicinity of the transition.
}
\label{fig:theomelt}  
\end{figure*}

Figure \ref{fig:theomelt} summarises the main results of the
theoretical analysis by showing the variation versus
temperature of the integrated intensities and
widths of the Bragg peaks in scan 1 and 2, as they are predicted by
the model. As expected the intensity shows a sharp drop near the
denaturation transition. It reflects the openings of the base pairs
that break the clusters of stacked pairs giving rise to the
investigated Bragg
peak and therefore reduces the number of scattering
sites. As a result the integrated intensity of the peak almost
provides a quantitative measure of the helix fraction of DNA because,
as shown by Eq.~(\ref{eq:sn1bragg}), for sufficiently large clusters,
the structure factor is proportional to the number of sites in a
cluster. Moreover Fig.~\ref{fig:theomelt}-c shows that the size of the
clusters significantly drops only in the last stage of the
denaturation. 

\smallskip
The width of the Bragg peak provides the spatial
information that standard observations of DNA denaturation cannot
give. It is strongly sensitive to the distribution of the sizes 
of the diffracting clusters. The drop of the average cluster size in
the last stage of the transition (Fig.~\ref{fig:theomelt}-c) is
reflected in the large increase of the width of the Bragg peak
predicted by the theory in the high temperature range
(Fig.~\ref{fig:theomelt}-d). For scan 2, 
with a nonzero transverse component $Q_{\perp}$ of the scattering vector,
the width of the peak is also affected by the transverse structural
disorder due to the effect of the sequence (variables $\eta_j$) and by
their correlations with the longitudinal structural disorder
(variables $\lambda_j$), measured by the coefficient $\chi$ in
Eq.~(\ref{eq:Gaussvar}).  The statistical properties of
$\lambda_j$ and $\eta_j$ have been obtained by conformational analysis
\cite{Lavery} but their correlations have not been determined. We show
results with $\chi = 0$ (no correlation) and $\chi = 0.35$
corresponding to moderate correlations. Moreover scan 2, is also
probing the transverse fluctuations of the bases prior to opening. The
inset in Fig.~\ref{fig:theomelt}-d shows that, in the vicinity of the
transition, these fluctuations are expected to cause an extra increase
of the width of the Bragg peak, for scan 2 only. However this effect
is small because the Bragg peak is only generated
by the closed sections of the DNA molecules, where the fluctuations
are therefore limited to small amplitude motions of the bases.

\begin{figure}[h]
\includegraphics[width=9.0cm]{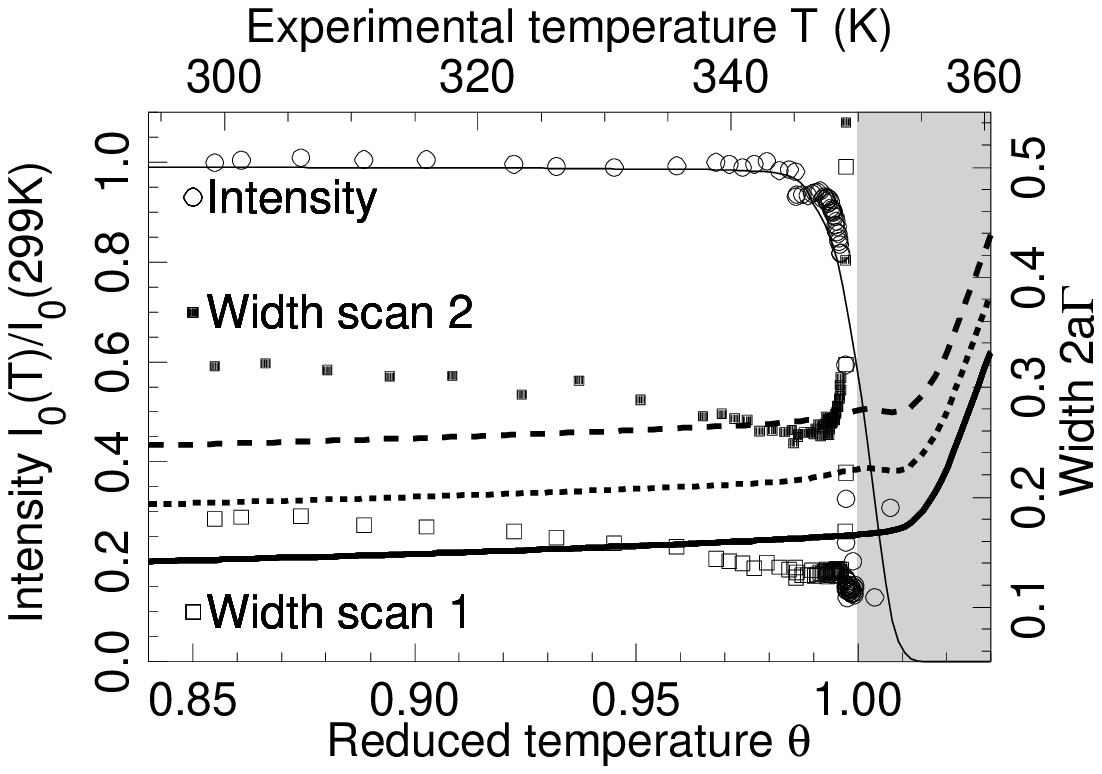}
  \caption{Comparison between theory and experiment. The points are
    the experimental results while the curves plot the theoretical
    results. A reduced temperature $\theta = T/T_c$ is used, where
    $T_c$ is temperature where 50\% of the bases are open.
    The circles show
  the integrated intensity of the Bragg peak, rescaled to 1 at low
  temperature. After rescaling, the results for scans 1 and 2 
  exactly follow  the same curve. The thin line is the calculated
  integrated intensity. The open squares show the experimental
  width of scan 1 and the thick full line is the theoretical value for
  this width. The filled squares show the width of scan 2. The
  dotted line shows the theoretical width of scan 2 calculated 
  by assuming that the longitudinal and transverse structural disorder
  due to the sequence, determined by conformational analysis
  \cite{Lavery}, are uncorrelated ($\chi = 0$) while 
 the dash line is the theoretical
  width of scan 2 calculated by assuming a partial correlation
  between the longitudinal and transverse structural disorder 
($\chi = 0.35$).
The shaded area of the plot shows the temperature range in which the
experimental observations are hindered by the collapse of the sample
film.
The figure is identical to Fig.~3 \cite{PRL}.}
  \label{fig:comparison}
\end{figure}

\section{Discussion}
\label{sec:discussion}

To allow a quantitative comparison of the experimental and theoretical
results we plot the data as a function of
a reduced temperature $\theta = T/T_c$, where $T_c$ is the
temperature where 50\% of the bases are open. This is done in
Fig.~\ref{fig:comparison}. To eliminate the experimental
factor associated to the apparatus, in this figure the experimental
intensity of the Bragg peak has been rescaled to $1$ in the low
$\theta$ limit, which is also the limit of the theoretical intensity
in the low temperature range. 
The widths have been multiplied by the mean separation of the base
pairs, $a$, to create dimensionless variables. Therefore a
quantitative comparison between theoretical and experimental
widths is possible and
it does not involve any arbitrary factor.

\smallskip
Let us first examine the integrated intensity of the Bragg peak. Both
the experimental data and the theoretical curve show an intensity
which stays almost constant up to temperatures very close $T_c$, 
$\theta \approx 0.97$. As previously noted this reflects the
very low fraction of open base pairs until the vicinity of the
transition is reached, in agreement with the theoretical
result of Fig.~\ref{fig:theomelt}-a. 
In the early stage of the transition the
theoretical curve follows the experimental decay of the
intensity of the peak, but in the immediate vicinity of $T_c$
($\theta \approx 1$) the experimental intensity shows an almost
discontinuous drop, while the theoretical curve has a narrow but
smooth decay. This discrepancy is a sign that another phenomenon, not
included in the theoretical description happens. As discussed in
Sec.~\ref{sec:experimental} the optical observations of a sample
during heating indicate that, at high temperature, the film itself shows
an irreversible ``collapse'' characterised by a disorganisation of the
oriented fibre structure. Although this is not described by the model,
the theoretical analysis shines some light on this phenomenon. Figure
\ref{fig:theomelt}-c shows that, above $T_c$, 
the length of the denatured regions
grows very quickly with temperature, showing almost a divergence. As
the single strands are very flexible they can gain a lot of entropy by
fully losing their initial orientation so that the remaining
double-helix segments are embedded in a liquid-like medium of
entangled single strands which quickly becomes the dominant phase in
the sample. The rapid growth of the size of the open fragments gives a
lot of freedom to the rigid double-helix segments allowing them 
to lose their
spatial orientation, which causes the sharp drop of the intensity of the
Bragg peak. At higher temperatures the sample is no longer a good
approximation of a one-dimensional crystal, but is more an ensemble
of disoriented, short length, DNA This temperature
range corresponds to the shaded area in Figs.~\ref{fig:theomelt} and
\ref{fig:comparison}. It is interesting to notice that, 
when this ``collapse'' of
the film occurs, the theory predicts that the size of the closed
segments is still large, of the order of $80 - 100$ base pairs. This
is indirectly confirmed by the electrophoresis analysis of the length
of the DNA fragments in the film before and after heating which
indicate that, after film melting, the DNA molecules that were more
than $20000$ base-pair long at low temperature are chopped into
segments of a few hundreds of base pairs. This breaking can be
understood by the high stress concentration that occurs at the end of
the rigid fragments, linked to each other by the flexible single
strands, when they rotate as the film melts. It is therefore not
surprising to detect DNA fragments which have a length of the order of
the size of the closed clusters.

\bigskip
Let us now examine the width of the Bragg peak. For scan 1
the calculation of the width does not involve any
free parameter once the model has been calibrated to match the
denaturation curve of DNA. The other parameters entering in the
structure factor calculation (Eqs.~(\ref{eq:sn}) and (\ref{eq:sq2}) )
are derived from the structure of DNA \cite{Lavery} and its 
sound velocity measured along the helix axis \cite{DNABrill}. For scan
1 ($Q_{\perp} 
= 0)$, the
calculation gives a result which is in good agreement with
experiments (Fig.~\ref{fig:comparison}), 
although the experimental results are probably affected
by some annealing of the sample causing a slight decay of
the width of the peak (Fig.~\ref{fig:heat-cool}) that the theory
does not describe. In spite of
this limitation two points emerge from the comparison between theory
and experiments. First the structural data which enter in the
calculation of the width for scan 1, and particularly the fluctuations
of the base-pair distances along the helix, measured by $\langle
\lambda^2 \rangle$, corresponding to a standard deviation of
$0.18\;$\AA, 
are here tested on a large scale since the width of the diffraction
peak involves an average over the  billions of base pairs of the
sample. The discrepancy
of less than 15\% between the calculated and experimental widths
indicates that the results of the conformational analysis \cite{Lavery} are
accurate. Second, for scan 1, the width of the peak is remarkably
constant until $T_c$ and the collapse of the film.
This indicates that the base pair openings, which start to
be very significant at $\theta = T/T_c = 0.98$ do not cause a sharp
decrease of the size of the diffracting clusters until the
denaturation has occurred. This is what the theoretical model indicates
(Fig~\ref{fig:theomelt}-c). Clusters of about 100 base pairs remain
intact well within the denaturation region and denature as a whole, this
being allowed by the surrounding open bubbles in the last stage of
the denaturation.
This provides a good test of the statistical physics description of
DNA that we use, validating the model beyond its ability to predict melting
curves.

\smallskip
Contrary to scan 1 the calculation of the the width of the Bragg peak
in scan 2, with a transverse component $Q_{\perp}$ of the scattering vector,
involves an unknown parameter, the coefficient $\chi$ that measures
the correlation between the longitudinal and transverse structural
disorder due to the sequence (Eq.~\ref{eq:Gaussvar}).  
Figure \ref{fig:comparison} shows that,
if we ignore this correlation by setting $\chi = 0$, the theoretical
value is about 30\% lower than the experimental width. Setting $\chi =
0.35$, i.e.\ a moderate correlation, we get a theoretical width which
matches the experimental value for $\theta \approx 0.97$ which we take
as a temperature where the sample is ``annealed''.
The results suggest that neutron
scattering could be used to probe this structural property of DNA and
it would be interesting to test this result by conformational
analysis.
\smallskip

In the range $0.99 < \theta <
1$ the experiment detects a significant rise of the width of scan
2 which is not shown by the theoretical curve.
According to Eq.~(\ref{eq:sn}), the transverse fluctuations of the
base pairs, which become large because they start to open, bring an
extra contribution to the width through a growth of 
$\langle \xi^2 \rangle$. 
This contribution is visible on the inset of Fig.~\ref{fig:theomelt},
but, as discussed in Sec.~\ref{sec:theory}, this effect has to be
small since only fluctuations in the closed clusters  of base pairs
can contribute to the shape of the Bragg peak. This is not enough to
account for the observed increase of the width of scan 2 below the
transition. However there is another contribution to the width which
is not included in the structure factor calculation, it is the
misalignment of the molecules. It is very likely that the
collapse of the film is preceded by increased fluctuations
in the orientation of the helix fragments. This should have a strong
influence on the width of the Bragg peak in scan 2. For instance
orientational fluctuations
of $10$ degrees which change the projection of the base pair distance
on $Q_{\parallel}$ by less than 2\% lead to a projection of 17\% of this
distance along $Q_{\perp}$. Only a theory of the collective effects leading
to the melting of the film could properly account for this effect.

\section{Conclusion}
\label{sec:conclusion}

In conclusion, we have shown that neutron scattering can be used to
monitor the thermal denaturation of DNA, providing the spatial
information that other methods cannot measure. By focusing the study
on the Bragg peak which is associated to the base pair stacking we
can obtain accurate results which are not limited by the fibre nature
of the samples. The width of the Bragg peak 
can be described by a simple nonlinear model for
DNA at the scale of base pairs, thus providing further validation of
this model which had already proved able to predict complex DNA
denaturation curves with a small number of parameters. Moreover, by
selecting a scattering vector which is not parallel to the axis of the
DNA helix, the shape of the Bragg peak is also sensitive to the
transverse fluctuations of the base pairs, that this dynamical model
can calculate. This should allow further comparison between theory and
experiments by investigating not only the opening of the base pairs,
but also their large scale fluctuations, important in many biological
processes. This aspect could not be investigated in the
present experiments due to the collapse of the fibre structure of the
sample.

\begin{acknowledgments}

We want to thank Dr.\ Monica Jimenez-Ruiz (Institut Laue Langevin) 
and the IN8 and IN3 instrument teams for helpful assistance. We also
thank Emmanuel Andr\'e for technical help for the optical microscopy
observations. 
Part of this work has been supported by the program Accueil-Pro of
R\'egion Rh\^one-Alpes.
\end{acknowledgments}


\begin{thebibliography}{0}%
\makeatletter
\providecommand \@ifxundefined [1]{%
 \@ifx{#1\undefined}
}%
\providecommand \@ifnum [1]{%
 \ifnum #1\expandafter \@firstoftwo
 \else \expandafter \@secondoftwo
 \fi
}%
\providecommand \@ifx [1]{%
 \ifx #1\expandafter \@firstoftwo
 \else \expandafter \@secondoftwo
 \fi
}%
\providecommand \natexlab [1]{#1}%
\providecommand \enquote  [1]{``#1''}%
\providecommand \bibnamefont  [1]{#1}%
\providecommand \bibfnamefont [1]{#1}%
\providecommand \citenamefont [1]{#1}%
\providecommand \href@noop [0]{\@secondoftwo}%
\providecommand \href [0]{\begingroup \@sanitize@url \@href}%
\providecommand \@href[1]{\@@startlink{#1}\@@href}%
\providecommand \@@href[1]{\endgroup#1\@@endlink}%
\providecommand \@sanitize@url [0]{\catcode `\\12\catcode `\$12\catcode
  `\&12\catcode `\#12\catcode `\^12\catcode `\_12\catcode `\%12\relax}%
\providecommand \@@startlink[1]{}%
\providecommand \@@endlink[0]{}%
\providecommand \url  [0]{\begingroup\@sanitize@url \@url }%
\providecommand \@url [1]{\endgroup\@href {#1}{\urlprefix }}%
\providecommand \urlprefix  [0]{URL }%
\providecommand \Eprint [0]{\href }%
\@ifxundefined \urlstyle {%
  \providecommand \doi  [0]{\begingroup \@sanitize@url \@doi}%
  \providecommand \@doi [1]{\endgroup \@@startlink {\doibase
  #1}doi:\discretionary {}{}{}#1\@@endlink }%
}{%
  \providecommand \doi  [0]{doi:\discretionary{}{}{}\begingroup
  \urlstyle{rm}\Url }%
}%
\providecommand \doibase [0]{http://dx.doi.org/}%
\providecommand \Doi [0]{\begingroup \@sanitize@url \@Doi }%
\providecommand \@Doi  [1]{\endgroup\@@startlink{\doibase#1}\@@Doi}%
\providecommand \@@Doi [1]{#1\@@endlink}%
\providecommand \selectlanguage [0]{\@gobble}%
\providecommand \bibinfo  [0]{\@secondoftwo}%
\providecommand \bibfield  [0]{\@secondoftwo}%
\providecommand \translation [1]{[#1]}%
\providecommand \BibitemOpen [0]{}%
\providecommand \bibitemStop [0]{}%
\providecommand \bibitemNoStop [0]{.\EOS\space}%
\providecommand \EOS [0]{\spacefactor3000\relax}%
\providecommand \BibitemShut  [1]{\csname bibitem#1\endcsname}%
\end{thebibliography}%


\begin{thebibliography}{xx}

\bibitem{Wilkins}
M.H.F. Wilkins, A.R. Stokes and H.R. Wilson, 
Nature {\bf 171} 738-740 (1953)

\bibitem{Franklin}
R.E. Franklin and R.G. Gosling, 
Nature {\bf 171} 740-741 (1953)

\bibitem{WatsonCrick}
J.D. Watson and F.H.C. Crick, 
Nature {\bf 171}, 737-738 (1953)

\bibitem{Gueron}
M. Gu\'eron, M. Kochoyan and J.-L. Leroy, Nature {\bf 328} 89-92 (1987)

\bibitem{Thomas}
R. Thomas,
{\it Biochimica et Biophysica Acta} {\bf 14}, 231-240 (1954)

\bibitem{Rice}
S.A. Rice and P. Doty,
{\it J. Am. Chem. Soc.} {\bf 79}, 3937-3947 (1957)

\bibitem{Wartell}
R.M. Wartell and A.S. Benight,
{\it Physics Reports} {\bf 126}, 67-107 (1985)

\bibitem{RefHRM}
C.T. Wittwer,
{\it Human Mutation} {\bf 30}, 857-859 (2009)

\bibitem{Poland}
D. Poland and H.A. Scheraga,
{\it J. Chem. Phys.} {\bf 45}, 1456-1463 (1966)
and
D. Poland and H.A. Scheraga,
{\it J. Chem. Phys.} {\bf 45}, 1464-1469 (1966)

\bibitem{PBD}
M. Peyrard and A.R. Bishop,
{\it Phys. Rev. Lett.} {\bf 62}, 2755-2758 (1989)
and T. Dauxois, M. Peyrard and A.R. Bishop,
{\it Phys. Rev. E} {\bf 47}, R44-R47 (1993)

\bibitem{NTh}
N. Theodorakopoulos,
{\it Phys. Rev. E} {\bf 82}, 021905-1-4 (2010);
  J. Nonlin. Math. Phys. (in press)

\bibitem{JOST}
D. Jost and R. Everaers,
{\it Genome wide application of DNA melting analysis}
J. Phys. Condens. Matter {\bf 21} 034108(14pp) (2009)

\bibitem{Cowley} R. A. Cowley, in \emph{Methods of Experimental
  Physics} vol. 23 part C. K. Sk\"old and D. L. Price (eds)
  (Academic, Orlando, 1987) pp. 1 - 68 

\bibitem{Ruprecht}
A. Rupprecht,
{\it Acta Chem. Scand.} {\bf 20}, 494-504  (1966) and
A. Rupprecht,
{\it Biotechnology and Bioengineering} {\bf XII}, 93-121 (1970)

\bibitem{Fuller}
W. Fuller, T. Forsyth and A. Mahendrasingam,
{\it Phil. Trans. R. Soc. Lond. B} {\bf 359}, 1237-1248 (2004)

\bibitem{PRL}
A. Wildes, N. Theodorakopoulos, J. Valle-Orero,
S. Cuesta-L\'opez, J.-L. Garden, and M. Peyrard,
PRL {\bf 106} 048101-1-4 (2011)

\bibitem{Grimm}
H. Grimm, H. Stiller, C.F. Majrzak, A. Rupprecht and U. Dahlborg,
{\it Phys. Rev. Lett.} {\bf 59}, 1780-1783 (1987)
 
\bibitem{Kamenetskii}
M.D. Frank-Kamenetskii,
{\it Biopolymers} {\bf 10}, 2623-2624 (1971)
and M.D. Frank-Kamennetskii,
{\it Physics Reports} {\bf 288}, 14-60 (1997)

\bibitem{Hohne}
G.W.H. H\"ohne, W.F. Hemminger and H.-J. Flammersheim,
{\it Differential Scanning Calorimetry} Springer (2010)

\bibitem{DNAglass}
J Valle-Orero, J.-L. Garden, J. Richard, A. Wildes, M. Peyrard,
{\it Glass transition behavior of melted DNA fibers: a comparative study with
PVac.} unpublished 

\bibitem{parax}
Xiang-Qi Mu,
{\it Acta Cryst.} {\bf A54}, 606-616 (1998)

\bibitem{Lavery}
R. Lavery, M. Moakher, J.H. Maddocks, D. Petkeviciute and K. Zakrzewska,
{\it Nucl. Acid Res.} {\bf  37}, 5917-5929  (2009)

\bibitem{DNABrill} 
M. Krisch, A. Mermet, H. Grimm, V.T. Forsyth and A. Ruprecht,
{\it Phys. Rev. E} {\bf 73}, 061909-1-10 (2006)


\bibitem{JPCM}
M. Peyrard, S. Cuesta L\'opez and D. Angelov,
{\it J. Phys. Condensed Matter} {\bf 21}, 034103-1-13 (2009)

\bibitem{Korolev}
N. I. Korolev, A.P. Vlasov and I.A. Kuznetsov,
{\it Biopolymers} {\bf 34}, 1275-1290 (1994)

\bibitem{Lindsay}
S.M. Lindsay, S.A. Lee, J.W. Powell, T. Weidlich, C. Demarco,
G.D. Lewen, J.N. Tao and A. Rupprecht,
{\it Biopolymers} {\bf 27}, 1015-1043 (1988)

\bibitem{vanerp}
T.S. van Erp, S. Cuesta L\'opez and M. Peyrard,
{\it Eur. Phys. J. E} {\bf 20}, 421-434 (2006)

\bibitem{zhang}
Zhang Yong-li, Zheng Wei-Mou, Liu Ji-Xing and Y.Z. Chen,
{\it Phys. Rev. E} {\bf 56}, 7100-7115 (1997)

\bibitem{NikosPRE}
N. Theodorakopoulos,
{\it Phy. Rev. E} {\bf 77}, 031919-1-8 (2008)

\bibitem{Datasheet}
SIGMA ALDRICH, product information sheet D1626: Deoxyribinucleic Acid
(DNA) sodium salt from salmon testes.

\bibitem{Genebank}
GenBank (http://www.ncbi.nlm.nih.gov/Genbank/)

\bibitem{Pabyssi}
G.N. Cohen, V. Barbe, D. Flament, M. Galperin, R. Heilig, O. Lecompte,
O. Poch, D. Prieur, J. Qu\'erellou, R. Ripp, J.C. Thierry, J. Van der Oost,
J. Weissenbach, Y. Zivanovic and P. Forterre,
{\it Mol. Microbiol.} {\bf 47}, 1495-1512 (2003)


\end{thebibliography}
\end{document}